\documentclass[12pt,journal,comsoc]{IEEEtran}  % Uncomment to create two-columns version

\usepackage[T1]{fontenc}% optional T1 font encoding
\usepackage{cite}
\usepackage{amsmath,amssymb,amsfonts}
\usepackage{algorithmic}
\usepackage{graphicx}
\usepackage{textcomp}
\usepackage{xcolor}
\usepackage{makecell}
\usepackage{multirow}
\usepackage{glossaries}
\usepackage{algorithm}
\usepackage{algorithmic}
\usepackage{hyperref}
\usepackage{bm}

\ifCLASSINFOpdf
\else
\fi
\usepackage{amsmath}
\usepackage[cmintegrals]{newtxmath}
\graphicspath{{fig/pdf/}}
\DeclareMathOperator*{\argmax}{arg\,max}

\def\BibTeX{{\rm B\kern-.05em{\sc i\kern-.025em b}\kern-.08em
    T\kern-.1667em\lower.7ex\hbox{E}\kern-.125emX}}
		
\newacronym{5gnr}{5GNR}{5G New Radio}
\newacronym{ack}{ACK}{acknowledgment}
\newacronym{api}{API}{application programming interface}
\newacronym{arq}{ARQ}{automatic repeat request}
\newacronym{bler}{BLER}{block error rate}
\newacronym{bs}{BS}{base station}
\newacronym{csma}{CSMA}{carrier-sense multiple access}
\newacronym{dl}{DL}{downlink}
\newacronym{fdm}{FDM}{frequency division multiplexing}
\newacronym{fdma}{FDMA}{frequency division multiple access}
\newacronym{harq}{HARQ}{hybrid automatic repeat request}
\newacronym{l2c}{L2C}{learning-to-communicate}
\newacronym{lbt}{LBT}{listen before talk}
\newacronym{mac}{MAC}{medium access control}
\newacronym{madrl}{MADRL}{multi-agent deep reinforcement learning}
\newacronym{marl}{MARL}{multi-agent reinforcement learning}
\newacronym{ota}{OTA}{over-the-air}
\newacronym{pdu}{PDU}{protocol data unit}
\newacronym{phy}{PHY}{physical layer}
\newacronym{pomdp}{POMDP}{partially observable Markov decision process}
\newacronym{rl}{RL}{reinforcement learning}
\newacronym{sdu}{SDU}{service data unit}
\newacronym{sg}{SG}{scheduling grant}
\newacronym{sr}{SR}{scheduling request}
\newacronym{tdm}{TDM}{time division multiplexing}
\newacronym{tdma}{TDMA}{time division multiple access}
\newacronym{ue}{UE}{user equipment}
\newacronym{ul}{UL}{Uulink}
\newacronym{urllc}{URLLC}{ultra reliable low latency communications}

\hypersetup{
pdftitle={Towards Joint Learning of Optimal MAC Signaling and Wireless Channel Access},
pdfsubject={Machine Learning for communications},
pdfauthor={Alvaro Valcarce, Jakob Hoydis},
pdfkeywords={wireless, channel access, signaling, multiagent reinforcement learning, learn to communicate}
}

\begin{document}
\title{Towards Joint Learning of Optimal MAC Signaling and Wireless Channel Access}

\author{Alvaro~Valcarce,~\IEEEmembership{Senior~Member,~IEEE,}
        and~Jakob~Hoydis,~\IEEEmembership{Senior~Member,~IEEE}
				\thanks{A. Valcarce and J. Hoydis are with Nokia Bell-Labs France, Route de Villejust, 91620 Nozay, France. Email: \{alvaro.valcarce\_rial, jakob.hoydis\}@nokia-bell-labs.com}% <-this % stops a space
}

\maketitle
\IEEEpeerreviewmaketitle

\begin{abstract}
Communication protocols are the languages used by network nodes.
Before a \gls{ue} exchanges data with a \gls{bs}, it must first negotiate the conditions and parameters for that transmission.
This negotiation is supported by signaling messages at all layers of the protocol stack.
Each year, the telecoms industry defines and standardizes these messages, which are designed by humans during lengthy technical (and often political) debates.
Following this standardization effort, the development phase begins, wherein the industry interprets and implements the resulting standards.
But is this massive development undertaking the only way to implement a given protocol?
% But is this the only way to develop a protocol? Could machines emerge their own signaling protocols automatically and without human intervention?
We address the question of whether radios can learn a pre-given target protocol as an intermediate step towards evolving their own.
Furthermore, we train cellular radios to emerge a channel access policy that performs optimally under the constraints of the target protocol.
We show that \gls{marl} and \gls{l2c} techniques achieve this goal with gains over expert systems.
Finally, we provide insight into the transferability of these results to scenarios never seen during training.

\end{abstract}

\begin{IEEEkeywords}
% wireless, channel access, signaling, multiagent reinforcement learning, learn to communicate
communication system signaling, learning systems, mobile communication
\end{IEEEkeywords}

% \fi
\IEEEpeerreviewmaketitle

\section{Introduction}
\IEEEPARstart{C}{urrent} \gls{mac} protocols obey fixed rules given by industrial standards (e.g., \cite{3GPP2020}).
These standards are designed by humans with competing commercial interests.
Despite the standardization process' high costs, the ensuing protocols are often ambiguous and not necessarily optimal for a given task.
This ambiguity increases the costs of testing, validation and implementation, specially in cross-vendor systems like cellular networks.
In this paper, we study if there might be an alternative approach to \gls{mac} protocol implementation based on reinforcement learning.

\begin{table*}[htbp]
	\caption{Channel access schemes, \gls{mac} protocols and the technologies that use them}
	\begin{center}
		\begin{tabular}{|c|c|c|c|c|c|c|c|}
			\hline
			\multicolumn{2}{|c|}{}& \multicolumn{6}{c|}{\textbf{Channel Access Schemes}} \\
			\cline{3-8} 
			\multicolumn{2}{|c|}{} & \multirow{3}{*}{\textit{\textbf{\acrshort{tdma}}}} & \multicolumn{3}{c|}{\textit{\textbf{\gls{fdma}}}} & \multirow{3}{*}{\textit{\textbf{CDMA}}} & \multirow{3}{*}{\textit{\textbf{Spatial}}}\\
			\cline{4-6} 
			\multicolumn{2}{|c|}{}& & \multirow{2}{*}{\textit{\textbf{Time-invariant}}} & \multicolumn{2}{c|}{\textit{\textbf{Time-variant}}} & & \\
			\cline{5-6}
			\multicolumn{2}{|c|}{}& & & \textit{\textbf{Non-OFDMA}} & \textit{\textbf{OFDMA}} & & \\
			\hline
			\multirow{4}{*}{\rotatebox[origin=c]{90}{\makecell{\textbf{\gls{mac}}\\ \textbf{protocols}}}} & \textit{\textbf{\makecell{Contention\\based}}}& \makecell{Aloha\\ \acrshort{csma}\\ Cognitive Radio}&\textit{N.A.} & \multicolumn{4}{c|}{Cognitive Radio} \\
			\cline{2-8} 
			 & \textit{\textbf{Coordinated}}& \makecell{GSM\\Token Ring\\Bluetooth}& \makecell{FM/AM radio\\DVB-T}& POTS & \makecell{LTE\\ \gls{5gnr}} &\makecell{3G\\GPS} & \makecell{Satcoms\\MIMO} \\
			\hline
		\end{tabular}
		\label{tabMASCP}
	\end{center}
\end{table*}

\begin{figure}
\centerline{\includegraphics[width=\columnwidth]{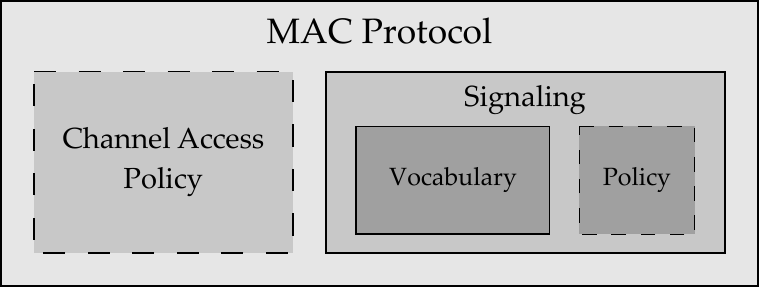}}  % Uncomment for Two-Columns final form
% \centerline{\includegraphics[width=4in]{MAC_protocol_architecture}}
\caption{\gls{mac} protocol constituent parts. This paper trains \glspl{ue} to jointly learn the policies highlighted as dashed blocks.}
\label{fig:MACProtocolArchitecture}
\end{figure}

To implement a \gls{mac} protocol for a wireless application, vendors must respect the standardized signaling (i.e., the structure of the \glspl{pdu}, the control procedures, etc) and the channel access policy (\gls{lbt}, \gls{fdma}, etc).
These are the main building blocks of a \gls{mac} protocol (see Fig \ref{fig:MACProtocolArchitecture}) and their implementation ultimately defines the performance that users will experience.
The signaling defines what control information is available to the radio nodes and when it is available.
In doing so, it constraints the actions that a channel access policy can take and puts an upper bound to its attainable performance.

% Factors influencing the design of a \gls{mac} protocol are network topology (cellular, mesh, device-to-device, etc), frequency range, deployment geometry (indoors-outdoors), etc.
% Even if the \glspl{phy} of two radios were inter-operable, \gls{mac} signaling differences would render co-existence/communication impossible.
% Furthermore, the limited set of messages standardized in industrial protocols constrain the type and amount of information that radios can share.
% These messages suffice to implement the channel access policies that industrial bodies have imagined.
% But they do not necessarily enable optimal channel access.

As wireless technologies evolve and new radio environments emerge (e.g., industrial IoT, beyond 6 GHz, etc), new protocols are needed.
The optimal \gls{mac} protocol may not be strictly contention-based or coordinated as shown in Table \ref{tabMASCP}.
For instance, \gls{5gnr} supports grant-free scheduling by means of a Configured Grant to support \gls{urllc}.
But are this acces policy and its associated signaling jointly optimal?
We are interested in the question of which \gls{mac} protocol is optimal for a given use case and whether it can be learned through experience.

\subsection{Related work}
Current research on the problem of emerging \gls{mac} protocols with machine learning focuses mainly on new channel access policies.
Within this body of research, contention-based policies dominate (see \cite{Yu2018}, \cite{Yu2019}, \cite{Destounis2019} or \cite{Yu2020}), although work on coordinated protocols also exists (e.g., \cite{Naderializadeh2019}, \cite{Naderializadeh2020}, \cite{AL-Tam2020}).
Other approaches such as \cite{Pasandi2020} propose learning to enable/disable existing \gls{mac} features.

Given a channel access scheme (e.g., \gls{tdma}), \cite{Yu2018} asks whether an agent can learn a channel access policy in an environment where other agents use heterogeneous policies (e.g., q-ALOHA, etc).
Whereas this is interesting, it solely focuses on contention-based access and it ignores the signaling needed to support it.
Instead, we focus on the more ambitious goal of jointly emerging a channel access policy and its associated signaling.

Dynamic spectrum sharing is a similar problem, where high performance has recently been achieved (see \cite{Bowyer2019}, \cite{Wong2020}) by subdividing the task into the smaller sub-problems of channel selection, admission control and scheduling.
However, these studies focus exclusively on maximizing channel-access efficiency under the constraints of a pre-given signaling.
None focuses on jointly optimizing the control signaling and channel access policy.

The field of Emergent Communication has been growing since 2016 \cite{Foerster2016}.
Advances in deep learning have enabled the application of \gls{marl} to the study of how languages emerge and to teaching natural languages to artificial agents (see \cite{Lowe2019} or \cite{Lowe2020}).
Due to the multi-agent nature of cellular networks and to the control-plane/user-plane traffic separation, these techniques generalize well to the development of machine-type languages (i.e., signaling).
In cellular systems, we interpret the \gls{mac} signaling as the language spoken by \glspl{ue} and the \gls{bs} to coordinate while pursuing the goal of delivering traffic across a network.

\subsection{Contribution}
For the reasons mentioned above, we believe that machine-learned \gls{mac} protocols have the potential to outperform their human-built counterparts in certain scenarios.
This idea has been recently used for emerging new digital modulations (see, e.g., \cite{AitAoudia2019}).
Protocols built this way are freed from human intuitions and may be able to optimize control-plane traffic and channel access in yet unseen ways.

The first step towards this goal is to train an intelligent software agent to learn an existing \gls{mac} protocol.
Our agents are trained tabula rasa with no previous knowledge or logic about the target protocol.
We show that this is possible in a simplified wireless scenario with \gls{marl}, self-play and tabular learning techniques, and lay the groundwork for scaling this further with deep learning.
In addition, we measure the influence of signaling on the achievable channel access performance.
Finally, we present results on the extension of the learned signaling and channel access policy to other scenarios.
\newline

The rest of this paper is organized as follows.
Section \ref{ProblemDefinition} formalizes the concepts of channel access, protocol and signaling.
It then formulates the joint learning problem and defines the research target.
Section \ref{sec:methods} describes the multi-agent learning framework.
Section \ref{sec:results} illustrates the achieved performance, provides an example of the learned signaling and discusses the transferability of these results.
A final discussion is provided in section \ref{sec:conclusions}.

\section{Problem definition}\label{ProblemDefinition}

\subsection{Definitions}
\begin{figure}
\centerline{\includegraphics[width=3in]{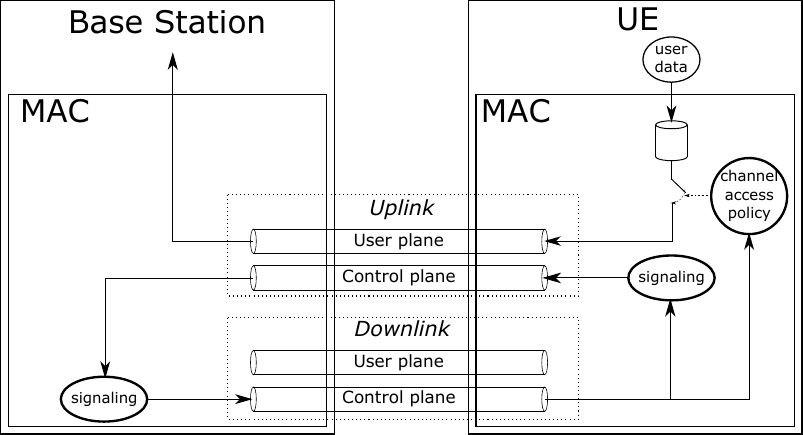}}  % Uncomment for Two-Columns final form
% \centerline{\includegraphics[width=5in]{UserControlPlanes}}
\caption{\gls{mac} protocol abstraction with \gls{ul} data traffic only.}
\label{fig:UserControlPlanes}
\end{figure}

We distinguish the concepts of \emph{channel access scheme} and \emph{\gls{mac} protocol} as follows:
\begin{itemize}
	\item \textbf{Channel access scheme}: Method used by multiple radios to share a communication channel (e.g., a wireless medium). The channel access scheme is implemented and constrained by the \gls{phy}. Sample channel access schemes are \gls{fdm}, \gls{tdm}, etc.
	\item \textbf{\gls{mac} protocol}: Combination of a \emph{channel access policy} and a \emph{signaling} with which a channel access scheme is used. Sample channel access policies are \gls{lbt}, dynamic scheduling, etc. Signaling is the vocabulary and rules (i.e., signaling policy) used by radios to coordinate and is described by the \gls{pdu} structure, the subheaders, etc. The channel access policy decides when to send data through the user-plane pipe, and the signaling rules decide when to send what through the control-plane pipe (see Fig. \ref{fig:UserControlPlanes}).
\end{itemize}
Table \ref{tabMASCP} illustrates these definitions with examples of technologies that use these mechanisms.

\subsection{Target \texorpdfstring{\gls{mac}}{MAC} signaling}

A complete commercial \gls{mac} layer provides numerous services to the upper layers (channel access, multiplexing of \glspl{sdu}, \gls{harq}, etc).
Our goal is to replace the \gls{mac} layer in a mobile \gls{ue} by a learning agent that can perform all these functions and their associated signaling.
However for simplicity, this paper targets a leaner \gls{mac} layer with two main functions: wireless channel access, and \gls{arq}.
Let this agent be denoted as the \gls{mac} \emph{learner}.
This differs from an \emph{expert} \gls{mac} implemented through a traditional design-build-test-and-validate approach.
Several learners are then concurrently trained in a mobile cell to jointly learn a \emph{channel access policy} and the \gls{mac} \emph{signaling} needed to coordinate channel access with the \gls{bs}.
The \gls{bs} uses an expert \gls{mac} that is not learned.

\begin{figure}
\centerline{\includegraphics[width=\linewidth]{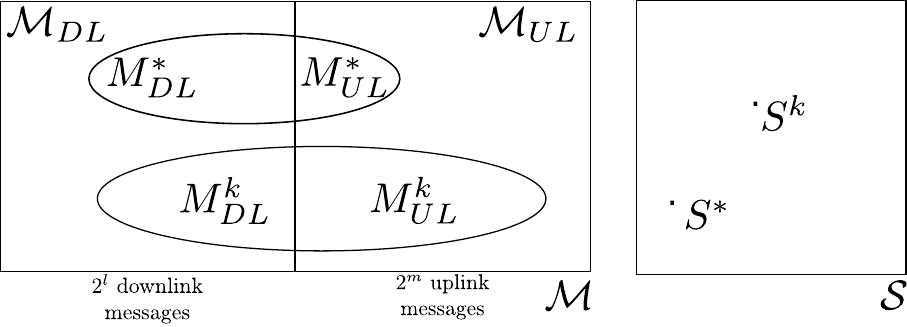}}  % Uncomment for final two-columns version
% \centerline{\includegraphics[width=4in]{Signaling}}
\caption{Left: Venn diagram of all possible signaling messages $\mathcal{M}=\mathcal{M}_{DL} \cup \mathcal{M}_{UL}$ of size $(l, m)$ bits. The set $M^*$ of messages for the optimal signaling $S^*$, as well as the set $M^k$ of messages for the $k^{th}$ arbitrary signaling $S^k$ are also shown.
Right: An optimum signaling $S^*$ and the $k^{th}$ arbitrary signaling $S^k$ within the abstract space $\mathcal{S}$ of all possible signaling (i.e., the signaling corpus).}
\label{fig:Signaling}
\end{figure}

Let $\mathcal{S}$ be the set of all possible \gls{mac} signaling that \glspl{ue} and a \gls{bs} may ever use to communicate (see Fig. \ref{fig:Signaling}).
We formalize a signaling as a vocabulary with \gls{dl} and \gls{ul} messages, plus mappings from observations to these messages.
Since different signalings are possible, let us denote the $k^{th}$ signaling $S^k\in\mathcal{S}$ as:
\begin{equation}
\begin{split}
S^k= [&M_{DL}^k, M_{UL}^k,\\ &O^{BS} \rightarrow M_{DL}^k,\\ &O^{UE} \rightarrow M_{UL}^k]
\end{split}
\end{equation}
where $M_{DL}^k\subseteq \mathcal{M}_{DL}$ and $M_{UL}^k\subseteq \mathcal{M}_{UL}$ are the sets of \gls{dl} and \gls{ul} messages of signaling $S^k$.
$O^{BS}$ and $O^{UE}$ are generic observations obtained at the \gls{bs} and the \gls{ue} respectively, and can include internal states, local measurements, etc.
In other words, a signaling defines the messages that can be exchanged and the rules under which they can be exchanged.
These rules give hence meaning to the messages.

In the above definitions, $|M_{DL}^k|$ and $|M_{UL}^k|$ are the sizes of the \gls{dl} and \gls{ul} signaling vocabularies, which implicitly define the amount of control data a single message can carry, i.e., the richness of the control vocabulary.
Messages from non-compositional protocols with larger vocabularies can therefore feed more control information to the channel access policy. Although this comes at the expense of a higher signaling overhead, the richer context available to the radio nodes can enable more sophisticated algorithms for channel access.
The size of the signaling vocabulary is hence an important hyper-parameter in emergent protocols to balance the trade-off between channel access performance and signaling cost.

The mappings from observations to messages define when to send what.
A \gls{mac} signaling policy $\pi_S$ describes one possible way of implementing this mapping.
This \emph{signaling policy} shall not be confused with the \emph{channel access policy}, which describes when and how to transmit data.

The \gls{bs} is an expert system with full knowledge of a standard \gls{mac} signaling $S^{BS}\in\mathcal{S}$.
Our first objective is to enable the \glspl{ue} to communicate with the \gls{bs} by learning to understand and use its signaling.
Out of the many signaling \glspl{ue} could learn, $S^{BS}$ is the learning target.
Note that $S^{BS}$ is not necessarily the optimal signaling $S^{*}\in\mathcal{S}$, which depends on a chosen metric, such as throughput, latency, etc.

The ideas presented here are generalizable to vocabularies of any size.
For simplicity, we have reduced the size of the target signaling vocabulary to the minimum number of messages needed to support both uncoordinated and coordinated \gls{mac} protocols.
In this paper, $S^{BS}$ has the following \gls{dl} messages that the \gls{mac} learners need to interpret:
\begin{itemize}
	\item \glspl{sg}
	\item \glspl{ack}.
\end{itemize}
Similarly, the \gls{ul} messages that the \gls{mac} learners need to learn to use are:
\begin{itemize}
	\item \glspl{sr}.
\end{itemize}
Other messages such as buffer status reports can be added to accommodate for larger vocabularies.

\subsection{Channel access policy}
The experiments described in this paper use a \gls{tdma} scheme, which was chosen for simplicity.
The learning framework proposed here is nevertheless equally applicable to other channel access schemes, such as those listed in Table \ref{tabMASCP}.

Unlike for the \gls{mac} signaling, we impose no a-priori known policy with which to use the access scheme.
This is to allow \glspl{ue} to explore the full spectrum of channel access policies between contention-based and coordinated.
For example, a channel access policy might follow a logic that ignores the available \gls{mac} signaling.
Other channel access policies may take it into consideration and implement a coordinated access scheme based on a sequence of \glspl{sr}, \glspl{sg} and \glspl{ack}.

Hence, the second objective of this research is for the \glspl{ue} to learn an \gls{ul} channel access policy $\pi_P$ that leverages the available signaling to perform optimally (according to some chosen channel access performance metric).
The channel access policy is denoted with the subscript $P$ to highlight that the \gls{mac} layer controls channel access by steering the underlying physical layer.
The \gls{mac} layer commands the \gls{phy} through an \gls{api} and is therefore constrained by the services it offers.
The \gls{mac} has no means of influencing the wireless shared channel without a \gls{phy} \gls{api} (e.g., \cite{Specification2019b}).
In our experiments, these services are limited to the in-order delivery of \gls{mac} \glspl{pdu} through a packet erasure channel.

\subsection{Channel model}
All learners share the same \gls{ul} frequency channel for transmitting their UL \gls{mac} \glspl{pdu} and they access this shared \gls{ul} data channel through their respective PHY layers.
From the viewpoint of the \gls{mac} learner, the PHY is thus considered part of the shared channel.
The channel accessed by each \gls{mac} learner is a \emph{packet erasure channel}, where a transmitted \gls{pdu} is lost with a certain \gls{bler}.
For simplicity, we assume that the \gls{bler} is the same for all data links and abstract away all other \gls{phy} features.
This comes without loss of generality to the higher-layer protocol analysis.

Since we want to study the effects of the control signaling on the performance of the shared data channel, we assume that the \gls{ul} and \gls{dl} control channels are error free, costless and dedicated to each user without any contention or collisions.

\subsection{\texorpdfstring{\gls{bs}}{BS} signaling policy}
Each time step, the \gls{bs} receives zero or more \glspl{sr} from the \glspl{ue}.
It then chooses one of the requesting \glspl{ue} at random and a \gls{sg} is sent in response.
An exception occurs if the \gls{ue} had made a successful data transmission concurrently with the \gls{sr}.
In this case, instead of an \gls{sg}, an \gls{ack} is sent to the \gls{ue} and another \gls{ue} is then scheduled at random.
This is because only one \gls{ue} can be scheduled each time slot.
More complex scheduling algorithms could be used here, but random scheduling has been chosen for simplicity.

\subsection{Multi-Agent Reinforcement Learning formulation}
\begin{figure}
\centerline{\includegraphics[width=\linewidth]{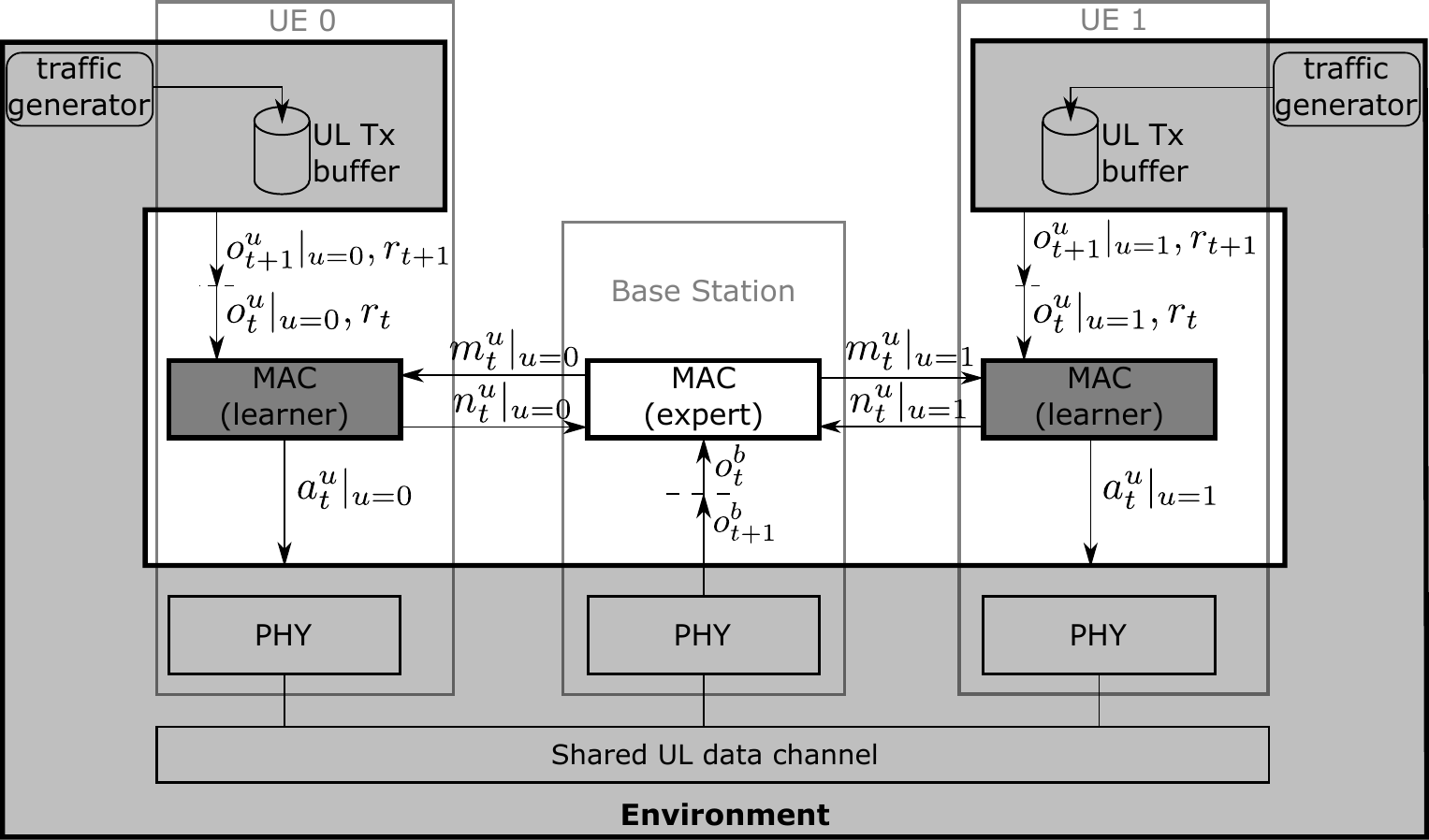}}  % Uncomment for two-column paper version
% \centerline{\includegraphics[width=13cm]{SystemModel_v3}}
\caption{System model with one \gls{bs} and two \glspl{ue}}
\label{fig:SystemModel}
\end{figure}

For any given \gls{ue}, the sequential decision-making nature of channel access lets us model it as a Markov Decision Process (MDP), which can then be solved using the tools of \gls{rl}.
Using \gls{rl} to train multiple simultaneous learners (i.e., \glspl{ue}) constitutes what is known as a \gls{marl} setup.
If the observations received by each learner differ from those of other learners, the problem becomes a \gls{pomdp}.

We have formulated this \gls{pomdp} as a cooperative Markov game (see \cite{Littman1994}), where all learners receive exactly the same reward from the environment.
Learners are trained cooperatively to maximize the sum $R$ of rewards:
\begin{equation}
R\triangleq\sum_{t=0}^{t_{max}} r_t\label{eq:performance}
\end{equation}
where $t_{max}$ is the maximum number of time steps in an episode.
This design decision reflects the objective of optimizing the performance of the whole cell, rather than that of any individual \gls{ue}.
An alternative approach that delivers different rewards to different \glspl{ue} depending on their radio conditions could perhaps yield higher network-wide performance.
However, this would require the design of multiple reward functions, which is known to be difficult (see \cite{WOLPERT2001}), and left to future investigations.

\subsubsection{Time dynamics}\label{TimeDynamics}
The \gls{marl} architecture is illustrated in Figure \ref{fig:SystemModel}.
$U$ is the set of all \gls{mac} learners.
Then, on each time step $t$, each \gls{mac} learner $u\in[0,|U|)$ invokes an action $a_t^u$ on the environment, and it receives a reward $r_{t+1}$ and an observation $o_{t+1}^u$.
The actions of all learners are aggregated into a joint action vector $\mathbf{a}_t$.
The environment then executes this joint action and delivers the same reward $r_{t+1}$ plus independent observations to all learners.
The \gls{bs} also receives a scalar observation $o_{t+1}^b$ following the execution of the joint learner action.

The environment studied in this paper demands that each \gls{mac} learner delivers a total of $P$ UL \gls{mac} \glspl{sdu} to the \gls{bs}.
We performed experiments with the following simple SDU traffic models:
\begin{itemize}
  \item \textbf{Full buffer start}: The UL Tx buffer is filled with $P$ SDUs at $t=0$.
	\item \textbf{Empty fuffer start}: The UL Tx buffer is empty at $t=0$. Then, it is filled with probability $0.5$ with one new SDU each time step until a maximum of $P$ SDUs have been generated.
\end{itemize}
The learners must also indicate awareness that the \glspl{sdu} have been successfully delivered by removing them from their UL transmit buffer.
An episode ends when the $P$ \glspl{sdu} of each and all of the $|U|$ \gls{mac} learners have successfully reached the \gls{bs} and the learners have removed them from their buffers.

\subsubsection{Observation space}
Each time step $t$, the environment delivers a scalar observation $o_t^u \in \mathcal{O^U}=[0, L]$ to each learner $u$ describing the number of \glspl{sdu} that remain in the learner's UL transmit buffer. Here, $L$ is the transmit buffer capacity and all \glspl{sdu} are presumed to be of the same size.
For example, the environment observation $o_t^u|_{u=5,t=2}=3$ indicates to learner $5$ that, at time $t=2$, three \glspl{sdu} are yet to be transmitted.

Similarly, at each time step the \gls{bs} receives a scalar observation $o_t^b \in \mathcal{O^B}=[0, |U|+1]$ from the environment.
This observation can take the following meanings:
\begin{itemize}
	\item $o_t^b=0$: \gls{ul} channel is idle
	\item $o_t^b\in[1,|U|]$: Collision-free UL transmission received from \gls{ue} $o_t^b-1$
	\item $o_t^b=|U|+1$: Collision in UL channel.
\end{itemize}
For example, the environment observation $o_t^b|_{t=4}=3$ indicates that, at time $t=4$, the \gls{bs} successfully received a \gls{mac} \gls{pdu} from \gls{ue} 2.

\subsubsection{Channel access action space}
Learners follow a channel access policy by executing actions $a_t^u\in\mathcal{A}_P$ every time step.
These actions have an effect on the environment by steering the physical layer and are $\mathcal{A}_P=\{0, 1, 2\}$, which are interpreted by the environment as follows:
\begin{itemize}
	\item $a_t^u=0$: Do nothing
	\item $a_t^u=1$: Transmit the oldest \gls{sdu} in the buffer. Invoking this action when the UL buffer is empty is equivalent to invoking $a_t^u=0$.
	\item $a_t^u=2$: Delete\footnote{A third action space with buffer-management actions could have also been defined. For simplicity, we only use this buffer related action and include it in $\mathcal{A}_P$} the oldest SDU in the buffer.
\end{itemize}
The channel access actions from all learners are aggregated into a joint action vector $\bm{a_t}=[a_t^0, a_t^1, ..., a_t^{|U|-1}]$, which is then executed on the environment at once. For example, invoking action $\bm{a_2}=[1, 2, 0]$ on the environment indicates that, at time $t=2$, \gls{ue} 0 attempts an \gls{sdu} transmission while \gls{ue} 1 deletes a \gls{sdu} from its buffer, and \gls{ue} 2 remains idle.

\subsubsection{Uplink signaling action space}
Following the approach introduced in \cite{Foerster2016}, in each time step, \gls{mac} learners can also select an \gls{ul} signaling action $n_t^u\in\mathcal{A}_S$.
This action maps to a message received by the \gls{bs} \gls{mac}, and it exerts no direct effect onto the environment.
These messages are thus transmitted through a dedicated \gls{ul} control channel that is separate from the shared \gls{ul} data channel modeled by the environment (see Fig. \ref{fig:UserControlPlanes}).
The \gls{ul} signaling actions available to the \gls{mac} learners are $\mathcal{M}_{UL}=\{0, 1\}$, which are interpreted by the \gls{bs} \gls{mac} expert as follows:
\begin{itemize}
	\item $n_t^u=0$: Null uplink message
	\item $n_t^u=1$: Scheduling Request.
\end{itemize}

\subsubsection{Downlink messages}
Finally, the \gls{bs} \gls{mac} expert can, each time step, select a \gls{dl} signaling action $m_t^u$ towards each \gls{mac} learner.
These messages have no direct effect on the environment and are transmitted through a dedicated \gls{dl} control channel.
The actions available to the \gls{bs} \gls{mac} expert are $\mathcal{M}_{DL}=\{0, 1, 2\}$ and have the following meanings:
\begin{itemize}
	\item $m_t^u=0$: Null downlink message
	\item $m_t^u=1$: Scheduling Grant for next time step
	\item $m_t^u=2$: ACK corresponding to the last SDU received in the uplink.
\end{itemize}
Note that the \gls{mac} learners are unaware of the meanings of these messages. They must learn them from experience.
Larger sizes of the \gls{dl} vocabulary $\mathcal{M}_{DL}$ would let the \gls{bs} communicate more complex schedules to the \glspl{ue} (e.g., by granting radio resources further into the future).
This would increase training duration significantly and is therefore out of the scope of this paper but left for future investigation.

\subsubsection{Learners' memory}
We endow each \gls{mac} learner with an internal memory $h_t^u\in\mathcal{H^U}$ to store the past history of observations, \gls{phy} and \gls{ul} signaling actions, as well as the DL messages received from the \gls{bs}.
$N$ denotes the size, in number of past time steps, of this internal memory, which at time step $t$ takes the form:
\begin{equation}
\begin{split}
h_t^u=[&m_{t-N}^u, a_{t-N}^u, n_{t-N}^u, o_{t-N}^u, ...,\\
&m_{t-1}^u, a_{t-1}^u, n_{t-1}^u, o_{t-1}^u].
\end{split}
\end{equation}
For example, if $N=1$, the internal memory $h_t$ at time $t$ contains only the observation, actions, and messages from time $t-1$.
The motivation for this memory is the need to disambiguate instantaneous observations that may seem equal to a given \gls{mac} learner, but emanate from the un-observed actions concurrently taken by other learners.
In short, memory addresses the problem of partial observability by the learners.
This is loosely based on the idea of \emph{fingerprinting} introduced in \cite{Foerster2017}.

The current state of the memory $h_t^u$ and the current observation $o_t^u$ constitute the two main inputs considered by the learners during action selection: $(a_t^u, n_t^u)=f(o_t^u,h_t^u)$.
For example, a learner $u$ with current memory state $h_t^u=[1, 0, 0, 1, 0, 0, 1, 1]$ and current observation $o_t^u=1$ describes a situation where the learner received a \gls{sg} in response to a previous \gls{sr} due to a non-empty Tx buffer.
The learner's current policy will then take this context into consideration for deciding the next channel access action and \gls{ul} signaling message (if any).

\subsubsection{Reward function}\label{sec:RewardFunction}
The environment delivers a reward of $r_t=-1$ on all time steps.
This motivates \gls{mac} learners to finish the episode as quickly as possible (i.e., in the smallest number of time steps).
The conditions for the termination of an episode were described in Section \ref{TimeDynamics}.

\section{Methods}\label{sec:methods}

\subsection{Tabular Q-learning}
\begin{figure}
\centerline{\includegraphics[width=2in]{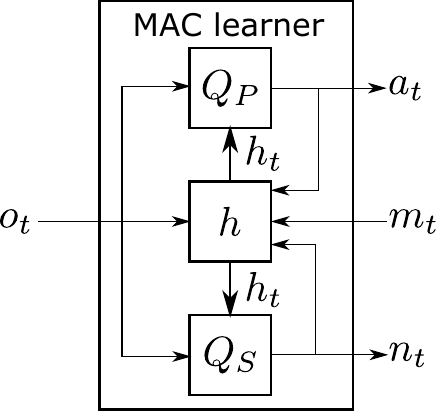}}
\caption{Internal structure of a \gls{mac} learner}
\label{fig:MAClearnerStructure}
\end{figure}

Given the definition of reward $r_t$ of section \ref{TimeDynamics}, the return $G_t$ at time $t$ is defined as (see \cite{Sutton2018}):
\begin{equation}
G_t\triangleq\sum_{k=0}^{\infty}\gamma^kr_{t+k+1}
\end{equation}
where $\gamma\in[0,1]$ is a discount factor.
The action-value function, under an arbitrary policy $\pi$, for a given observation-action pair $(o_t, a_t)$ is then defined as the expected return when that observation is made and the action taken:
\begin{equation}
Q_{\pi}(o_t, a_t)=E_{\pi}[G_t|a_t, o_t].
\end{equation}
Q-learning \cite{Watkins1989} is a well known technique for finding the optimal action-value function, regardless of the policy being followed.
It stores the value function $Q(o, a)$ in two-dimensional tables, whose cells are updated iteratively with:
\begin{equation}
\begin{split}
Q(o_t, &a_t) \leftarrow Q(o_t, a_t) +\\ &\alpha [r_{t+1} + \gamma \max_a Q(o_{t+1}, a) - Q(o_t, a_t)].
\end{split}
\end{equation}
Independently of how the table is initialized, $Q$ will converge to the optimal action-value function as long as sufficient exploration is provided.

Independent Q-learners are known to violate the Q-learning convergence requirements (see \cite{Tan93multi-agentreinforcement}). However, in practice, good results have been obtained when combining it with memory-capable agents and decentralized training.
These two techniques are essential to address the non-stationarity of multi-agent Q-learning problems and have been successfully applied in \cite{Foerster2016} on an \gls{l2c} setting. Our \gls{mac} learner architecture and training procedure are a tabular adaptation of the RIAL architecture described in \cite{Foerster2016}, whose success is one of the main motivations for using Q-learning here.

Each \gls{mac} learner is composed of two Q tables (see Fig. \ref{fig:MAClearnerStructure}), which contain action value estimates $Q_P$ for the physical-layer actions and action value estimates $Q_S$ for the signaling actions.
The reduced size of the problem described in Section \ref{ProblemDefinition} allows for the application of a traditional Q-learning approach based on Q tables (see \cite{Sutton2018}).
This is less data hungry than deep learning approaches, i.e., it needs less samples to converge and is therefore appropriate as a proof of concept.
However, tabular Q-learning scales poorly and is certainly not suitable in larger problems (e.g., $|U|>>2$, more SDUs, larger memories, etc.).
In such cases, deep learning approaches are needed (the tables in Fig. \ref{fig:MAClearnerStructure} can be easily replaced by deep neural networks).

Then, every time step $t$, each \gls{mac} learner follows two different policies $\pi_P$ and $\pi_S$ to act:
\begin{equation}
\begin{split}
\pi_P&: \mathcal{O^U} \times \mathcal{H^U} \longrightarrow \mathcal{A}_P\\
\pi_S&: \mathcal{O^U} \times \mathcal{H^U} \longrightarrow \mathcal{M}_{UL}.
\end{split}
\end{equation}
Although $a_t\in\mathcal{A}_P$ and $n_t\in\mathcal{M}_{UL}$ actions are chosen independently, both policies are synchronous due to the conditioning of both Q tables on the same current observation and state of the internal memory:
\begin{equation}
a_t = \argmax_a Q_P(a|o_t, h_t)
\end{equation}
\begin{equation}
n_t = \argmax_n Q_S(n|o_t, h_t).
\end{equation}

% Furthermore,  the \gls{ue}'s internal state at time $t$ depends on the downlink message $m_{t-1}$ received from \gls{bs} at time $t-1$. And in turn, $m_{t-1}$ is dependent on the signaling action chosen by the \gls{ue} at time $t-2$.

\subsection{Training procedure}
\gls{mac} learners are trained using self-play \cite{Tesauro1994}, which is known to reach a Nash equilibrium of the game when it converges.
This yields policies that can be copied to all \glspl{ue} and used during deployment.
Training is centralized, with one central copy of the $Q_P$ and $Q_S$ tables being updated regularly as experience from all \gls{mac} learners is collected.
In our experiments, we update the tables after each time step, although other schedules are possible.
Then, each time step, all learners choose their actions based on the same version of the trained tables (i.e., decentralized execution) combined with an $\epsilon$-greedy policy. The exploration probability $\epsilon$ is annealed between training episodes with $\epsilon_e=max(\epsilon_{e-1}*F_{\epsilon}^e, 0.01)$, where $\epsilon_0=1$, $F_{\epsilon}$ is the exploration decay factor, and $e>0$ is the training episode number.
Asymptotically, $\epsilon$-greedy guarantees that learners try all actions an infinite number of times. This ensures that $Q(o_t, a_t)$ converges to the optimal $Q^*(o_t, a_t)$. However, training is never run indefinitely, so the theoretical guarantee looses relevance and leaves the door open to alternative exploration methods such as e.g., optimistic initial action values (see \cite{Sutton2018}).
This \emph{centralized training with decentralized execution} approach has been chosen to address the non-stationarity pathology typical of setups with independent learners \cite{Hernandez-Leal2018}.
Deploying the same policies to all learners reduces the variance of observations during learning and helps convergence.
This training procedure can be performed in a server farm, where thousands of cross-vendor \glspl{ue} (cabled and/or \gls{ota}) would contribute with experiences to the training of a central \gls{ue} \gls{mac} model.
From a practical viewpoint, centralized training also spares \glspl{ue} from executing costly training algorithms. This way, most of the computational workload is shifted towards a central server. This avoids impacting the \glspl{ue}' battery performance and escapes constraints due to \gls{ue} mobility.
It is precisely this procedure which could replace the more traditional design-build-test-and-validate approach to \gls{mac} layer development in future protocol stacks.

Different \gls{bs} models can also be incorporated to the training testbed to increase the environment variance.
This can generalize the learned policies and improve performance when deployed in mobile networks different from the ones used during training.
Along these lines, zero-shot coordination techniques (see \cite{Hu2020}) could also help.

The system was trained for a fixed number $N_{tr}$ of consecutive training episodes, followed by a fixed number $N_{eval}$ of consecutive evaluation episodes.
This sequence of $N_{tr}+N_{eval}$ episodes is called a training session.
The Q tables were only updated during training episodes, but not during evaluation episodes.
The $\epsilon$-greedy policy with $\epsilon$ decay was followed only during training episodes.
Evaluation episodes follow strictly the learned policy and are thus free of exploration variance.
For each configuration (set of hyper-parameters), training sessions were repeated $N_{rep}$ times with different random seeds.
The convergence time grows with the problem size, specially with hyper-parameters $P$ and $|U|$. Fig. \ref{fig:training_curve_u2_p2_n4_lr0p3_vs_bler} illustrates how even a small scenario with $|U|=2$ \glspl{ue} and $P=2$ \glspl{sdu} requires several million training episodes to converge.

\section{Results}\label{sec:results}
\begin{table}[htbp]
\caption{Scenario settings and simulation hyper-parameters}
\begin{center}
\begin{tabular}{|c|c|c|}
\hline
\textbf{Symbol}& \textbf{Name} & \textbf{Value} \\
\hline
\gls{bler} & block error rate & $0, 10^{-4}, ..., 10^{-1}$ \\
\hline
SB & start buffer & full, empty \\
\hline
$P$ & number of \gls{mac} \glspl{sdu} & $1$, $2$ \\
\hline
$|U|$ & number of \glspl{ue} & $1$, $2$ \\
\hline
$t_{max}$ & maximum time steps per episode & $4, 8, 16, 32$ \\
\hline
$N_{tr}$ & number of training episodes & $2^{13}, 2^{14}, ...$ \\
\hline
$N_{eval}$ & number of evaluation episodes & 128 \\
\hline
$N_{rep}$ & training session repetitions & 4, 8 \\
\hline
$\gamma$ & discount factor & 1 \\
\hline
$F_{\epsilon}$ & exploration decay factor & $0.999991, 0.9999991$\\
\hline
$\alpha$ & learning rate & 0.3 \\
\hline
$N$ & memory length of \gls{mac} learner & $1, 2, 3, 4$ \\
\hline
\end{tabular}
\label{tabHyperParams}
\end{center}
\end{table}
The training procedure described above was tested in simulation, and its performance $R$ measured as the reward collected per episode (see \eqref{eq:performance}).
The optimal hyper-parameters have been chosen via grid search in the following discretized sets:
\begin{itemize}
	\item Discount factor: $[0, 0.5, 1]$
	\item Exploration decay factor: $0.99991, 0.999991, 0.9999991$
	\item Learning rate: $[0.05, 0.06, 0.1, 0.2, 0.3, 0.4]$
\end{itemize}
Table \ref{tabHyperParams} collects the best-performing parameters.

\subsection{Expert baseline}
\begin{algorithm}
\caption{Expert channel access policy}
\label{alg:ExpertChannelAccessPolicy}
\begin{algorithmic}
\REQUIRE $o_t, m_{t-1}, a_{t-1}$
\IF{$m_{t-1}=\gls{sg}$ \AND $o_t>0$ \AND $a_{t-1}\neq1$}  % If a \gls{sg} was received at $t-1$, there's at least one \gls{sdu} in the Tx buffer and no transmission was made at $t-1$
\STATE Transmit oldest \gls{sdu} in the Tx buffer ($a_t = 1$)
\ELSIF{$m_{t-1}=\gls{ack}$ \AND $o_t>0$}  % If an \gls{ack} was received at $t-1$ and there's at least one \gls{sdu} in the Tx buffer
\STATE Delete oldest \gls{sdu} in the Tx buffer ($a_t = 2$)
\ELSE
\STATE Do nothing ($a_t=0$)
\ENDIF
\end{algorithmic}
\end{algorithm}

\begin{algorithm}
\caption{Expert signaling policy}
\label{alg:ExpertSignalingPolicy}
\begin{algorithmic}
\REQUIRE $o_t, m_{t-1},$
\IF{$o_t > 0$}
\IF{$m_{t-1} = 0$ \OR ($o_t > 1$ \AND $m_{t-1} = \gls{ack}$)}
\STATE Send an \gls{sr} to the \gls{bs} ($n_t = 1$)
\ELSE
\STATE Do nothing ($n_t=0$)
\ENDIF
\ELSE
\STATE Do nothing ($n_t=0$)
\ENDIF
\end{algorithmic}
\end{algorithm}

For comparison purposes, the performance obtained by a population of expert (i.e., non-learner) \glspl{ue} is also shown.
The expert \gls{ue} has complete knowledge about the \gls{bs} signaling semantics and only transmits following the reception of an \gls{sg}.
Similarly, it only deletes an \gls{sdu} from its buffer following reception of an \gls{ack}.
These \glspl{ue} follow a fully coordinated channel access policy and do not profit from potential gains due to stochastic contention.
Pseudo code for the channel access and signaling policies implemented by the expert \gls{ue} is provided in the Algorithms \ref{alg:ExpertChannelAccessPolicy} and \ref{alg:ExpertSignalingPolicy}, respectively.

\subsection{Optimal policy}
The optimal policy $\pi^{*}=(\pi^{*}_P, \pi^{*}_S)$ for a single \gls{mac} learner that needs to transmit a single \gls{sdu} (i.e., $|U|=1$ and $P=1$) can be intuitively deducted and its performance modeled analytically.
Considering that the total reward that can be collected in a given episode depends on $t_{max}$, there may be different optimal policies for different values of this parameter.
Indeed, the optimal channel access policy for low $t_{max}$ consists in transmitting an \gls{sdu} at $t=0$ and immediately removing it from the Tx buffer in the next time step.
\glspl{ue} using this policy ignore the \glspl{ack} because on average, waiting for the ACK before removing the SDU takes too long.
Let this optimal policy be $\pi^{(1)}$ with expected performance:
\begin{equation}
R^{(1)}=b\cdot(2-t_{max})-2\label{eq_R1}
\end{equation}
where $b$ denotes the \gls{bler}.
When the \gls{sdu} transmission at $t=1$ succeeds, this leads to a sum reward of $-2$.
If the transmission fails, the UE will get, under policy $\pi^{(1)}$, the mininum reward of $-tmax$.
The value of $tmax$ for which this policy is optimal depends hence on the \gls{bler}.

\begin{figure}
\centerline{\includegraphics[width=\linewidth]{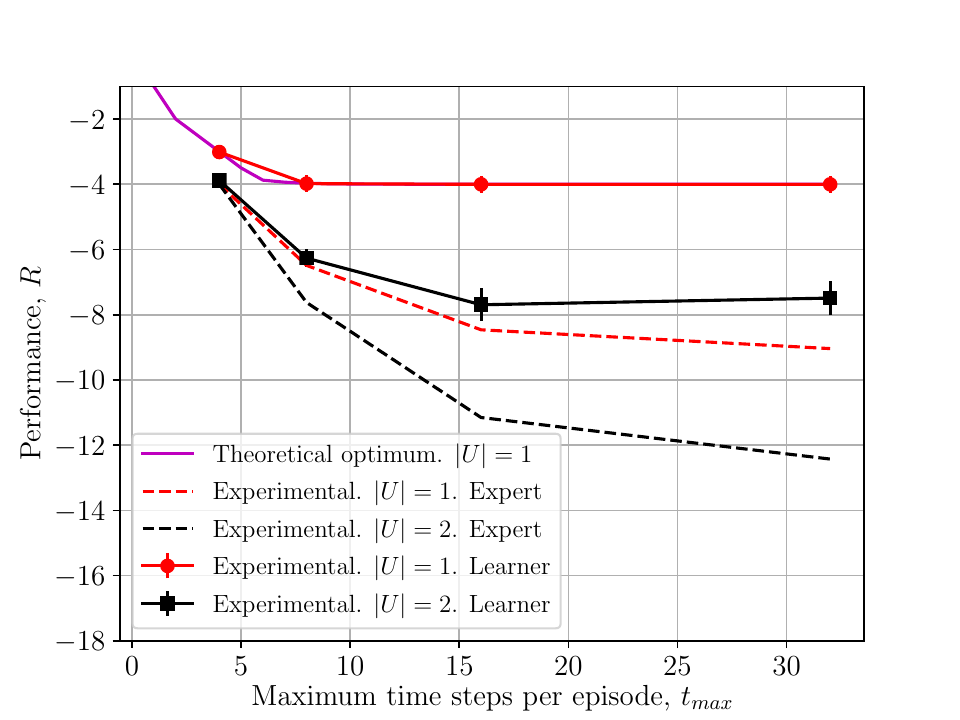}}
\caption{Average performance over eight independent training sessions. The performance is evaluated over $N_{eval}=128$ episodes with different seeds. The environment was configured with $P=1$, $BLER=0.5$ and start buffer full. \gls{mac} learners were trained with a learning rate of $0.05$. Learners on the $|U|=1$ environment were trained with an internal memory length of $N=1$ for $N_{tr}=2^{13}=8192$ episodes, while learners on the $|U|=2$ environment were trained with an internal memory length of $N=3$ for $N_{tr}=2^{18}=262144$ episodes. No exploration was used (i.e., $\epsilon=0$). The length of the error bars equal the standard error of the mean performance (calculated over the $8$ independent training sessions).}
\label{fig:Performance_u1}
\end{figure}

For higher $t_{max}$, the \gls{mac} learner is encouraged to wait for an \gls{ack} because the risk of ignoring it is too high due to the long episode duration.
Let this optimal policy for $|U|=1$ be $\pi^{(2)}$.
One can easily derive the expected performance analytically as:
\begin{align}
R^{(2)}= 
\begin{cases}
		-t_{max},& t_{max} < 4\\
		-(b+3),& t_{max} = 4\\
		\\
		\makecell[l]{(b-1) (3 + \sum\limits_{i=4}^{t_{max}-1}i b^{i-3})\\ - t_{max}\cdot b^{t_{max}-3}              }& t_{max} > 4
\end{cases}\label{eq_R2}.
\end{align}
The expected optimum performance in this scenario ($|U|=1$ and $P=1$) is shown in Fig. \ref{fig:Performance_u1} and is:
\begin{equation}
R^{*}=max(R^{(1)}, R^{(2)}).
\end{equation}

It is worth noting that the \gls{mac} learners' behavior is an artifact of the reward function.
The reward described in section \ref{sec:RewardFunction} encourages them to accomplish a task as fast as possible.
Alternatively, one could provide a reward of 1 (or 1 for each message) as soon as the messages have been successfully delivered and a reward of -1 if the maximum duration has passed.

Fig. \ref{fig:Performance_u1} shows how \gls{mac} learners use $\pi^{(1)}$ or $\pi^{(2)}$ depending on the maximum episode length.
Both of these policies ignore \glspl{sg} and are therefore useless in scenarios with more than $|U|=1$ \gls{ue}, where collisions impose the need for \gls{mac} learners to respect the scheduling allocation decided by the \gls{bs}.

Fig. \ref{fig:Performance_u1} also shows experimental results for the $|U|=2$ scenario to illustrate how performance scales with increasing numbers of \glspl{ue}.
Similarly to the single-\gls{ue} case, ACKs are largely ignored in the low $t_{max}$ regime when $|U|=2$ \glspl{ue} are present.
In this larger scenario, the vast size of the signaling solution space makes it difficult to find the optimal policy, much less a closed-form expression for its performance.
Nevertheless, a measure of gain can still be obtained by comparing the performance of the learned policies against that of a known expert (see Fig. \ref{fig:Performance_u1} for the gains above the expert).

\subsection{\texorpdfstring{\gls{bler}}{BLER} impact on \texorpdfstring{\gls{mac}}{MAC} training}
\begin{figure}
\centerline{\includegraphics[width=\linewidth]{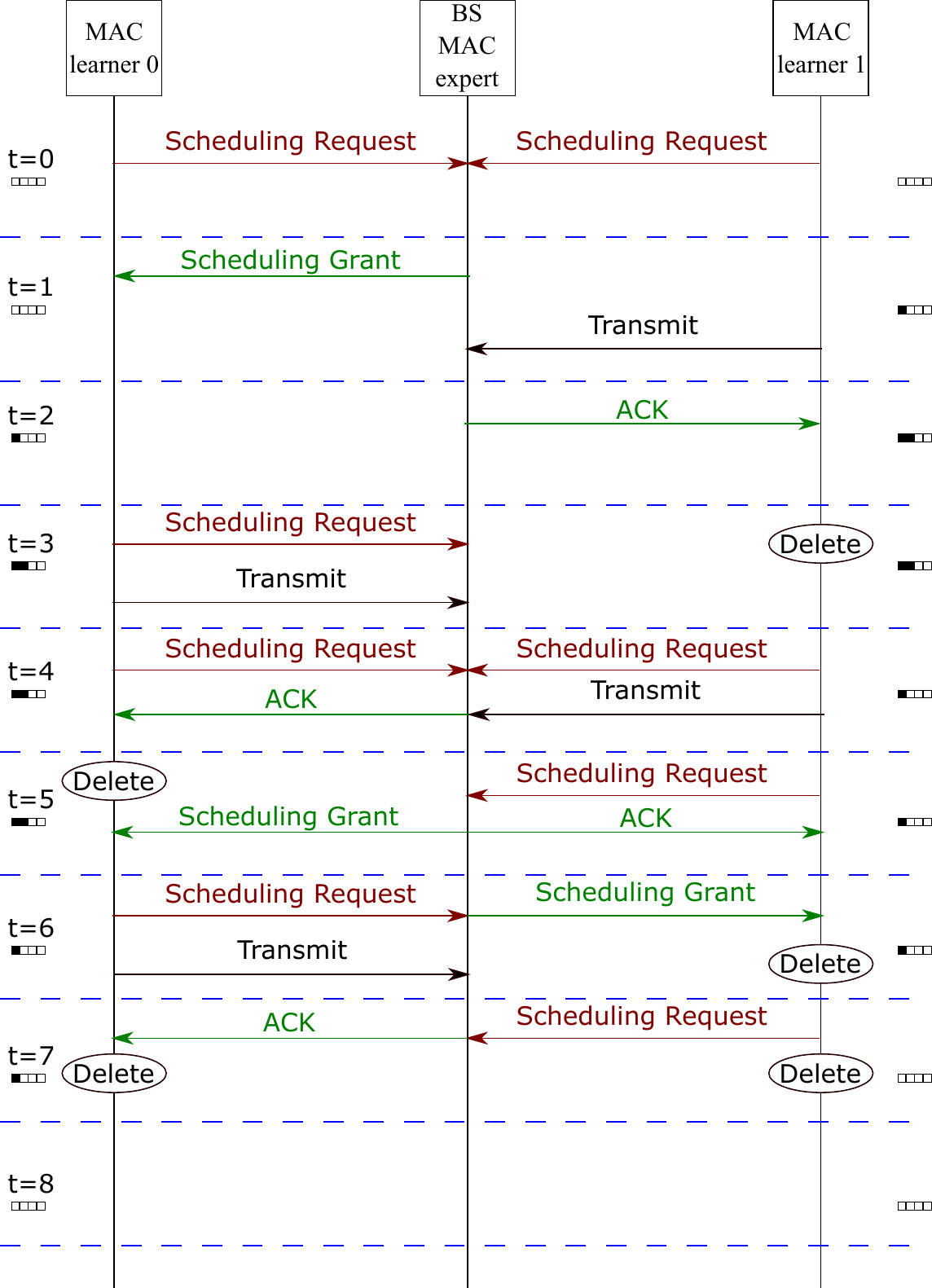}}  % Uncomment for two-columns version
% \centerline{\includegraphics[width=13cm]{MSC_exp867_Rep0_ValEp36.pdf}}
\caption{Best performing evaluation episode in a scenario with $|U|=2$, $P=2$, $N=4$, $t_{max}=32$, empty buffer start and $BLER=10^{-3}$. The small 4-cells grid depicts the UL Tx buffer, with one dark cell per available SDU.}
\label{fig:MSC}
\end{figure}

In the absence of block errors, \gls{mac} learners learn to ignore \glspl{ack}.
On the other hand, unexpected \gls{pdu} losses motivate \gls{mac} learners to interpret the \gls{dl} ACKs before deleting transmitted \glspl{sdu} from the Tx buffer.
This is illustrated in the learned \gls{mac} signaling trace of Fig. \ref{fig:MSC}, which shows the \gls{mac} learners removing \glspl{sdu} from the UL Tx buffer at time steps immediately following the reception of an ACK (deletions at $t=3$ and $t=6$ for \gls{mac} 1, and at $t=5$ for \gls{mac} 0).

Interestingly, \gls{mac} 0 also removed a SDU at $t=7$ before it had time to process the received \gls{ack}.
The reason is because, unlike in all other transmissions, transmission at $t=6$ was preceded by a \gls{sg} at $t=5$, which guarantees it to be collision-free.
In this scenario, \gls{bler} was so low that, on average, removing the \gls{sdu} from the buffer before waiting for an \gls{ack} yields a higher average reward.
This suggests that \gls{mac} protocols may perform better by skipping \glspl{ack} during dynamic scheduling in low \gls{bler} regimes.
This is clearly something not typically done in human-designed protocols.

\begin{figure}[t!]
\centerline{\includegraphics[width=\linewidth]{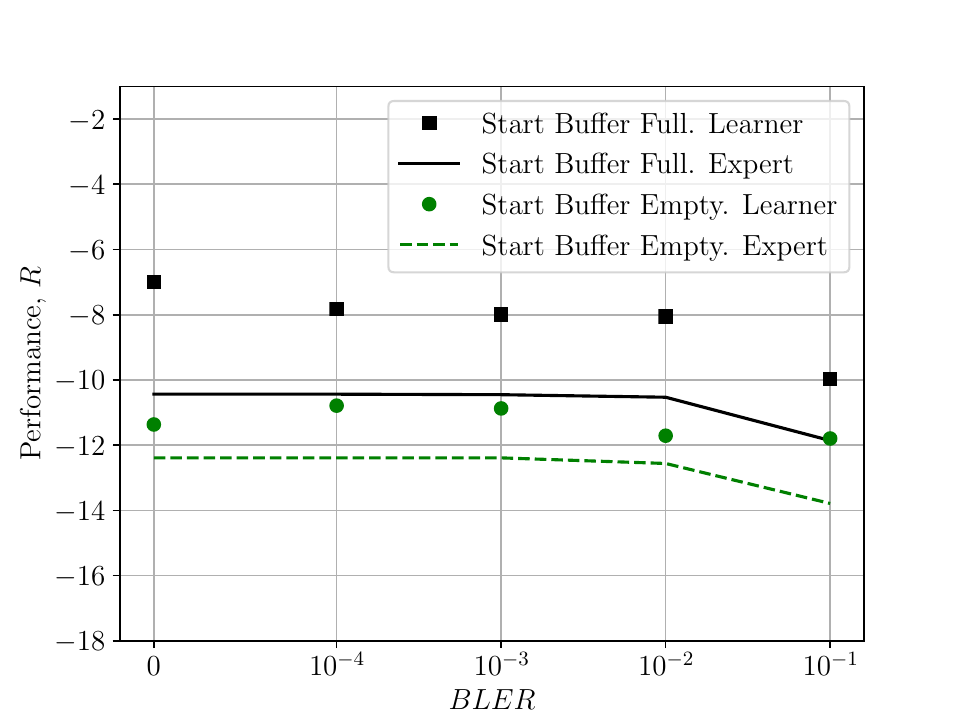}}
\caption{Best performance out of $4$ independent training sessions. The performance is evaluated over $N_{eval}=128$ episodes with different seeds. The environment was configured with $|U|=2$, $P=2$ and $t_{max}=32$. \gls{mac} learners were trained with learning rates between $0.1$ and $0.3$ and an internal memory length of $N=4$ . Start buffer full learners were trained for $N_{tr}=2^{20}\approx 1$ million episodes, while start buffer empty learners were trained for $N_{tr}\leq 4$ million episodes. $\epsilon$-greedy exploration with $\epsilon$ decay from 1 to 0.01 was used.}
\label{fig:Performance_u2_p2_N4}
\end{figure}

\begin{figure}
\centerline{\includegraphics[width=\linewidth]{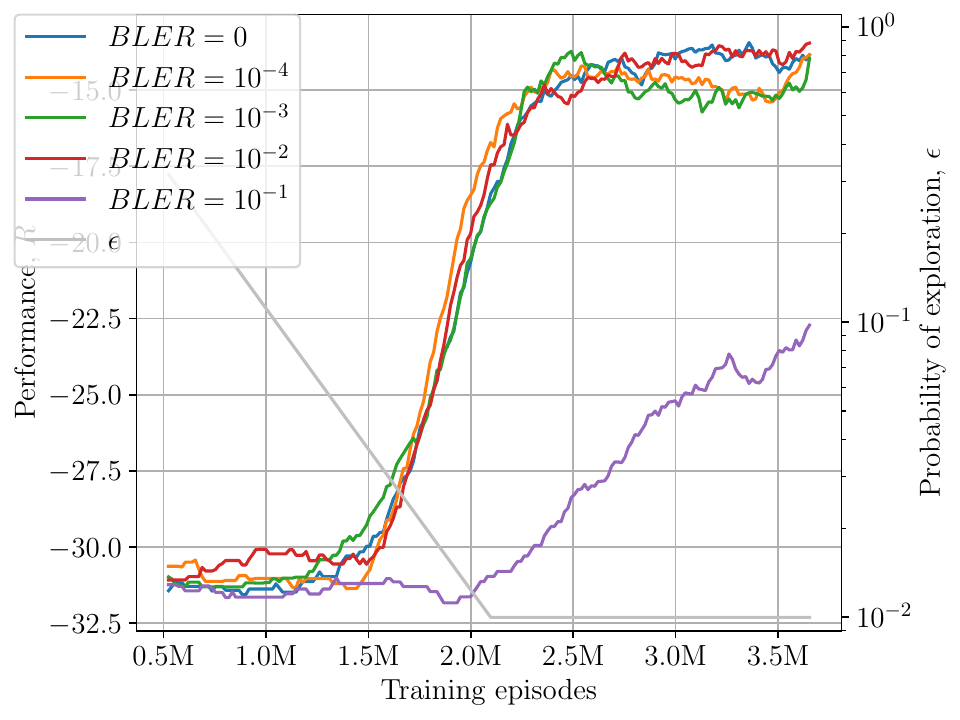}}
% \centerline{\includegraphics[width=11cm]{training_curve_u2_p2_n4_lr0p3_vs_bler.pdf}}  % For draft
\caption{Average learning progress (over $4$ independent training sessions) under various \gls{bler} conditions. The environment was configured with $|U|=2$, $P=2$, $t_{max}=32$ and empty start buffer. \gls{mac} learners were trained with a learning rate of $0.3$ and an internal memory length of $N=4$ for $N_{tr}=2^{22}\approx 4$ million episodes. $\epsilon$-greedy exploration with $\epsilon$ decay from 1 to 0.01 was used.}
\label{fig:training_curve_u2_p2_n4_lr0p3_vs_bler}
\end{figure}

As expected, having to wait for the ACKs before deleting \glspl{sdu} from the Tx buffer reduces performance, since episodes take longer to complete (see Fig. \ref{fig:Performance_u2_p2_N4}).
The presence of \gls{bler} in the data channel also slows down training due to a larger number of state transitions (see Fig. \ref{fig:training_curve_u2_p2_n4_lr0p3_vs_bler}).

\subsection{The importance of signaling}
A major question that arises in \textit{learning-to-communicate} settings is whether gains are due to either optimized action policies or better communication (i.e., signaling in our case), or both.
In our \gls{marl} formulation, \gls{mac} learners have two action spaces and are trained to learn two distinct policies (i.e., a channel access and a signaling policy).
Hence, in an extreme case, it is conceivable for the learners to ignore all DL signaling and learn an optimized channel access policy instead.
In this case, no \gls{mac} protocol signaling would be needed.

To address the previous question, we have calculated the Instantaneous Coordination ($IC$) metric proposed in \cite{Jaques2018}.
For the $u^{th}$ \gls{ue}, $IC^u$ is defined in this scenario as the mutual information between the \gls{bs}'s DL \gls{mac} messages received at time $t$ and the \gls{ue} \gls{mac}'s channel access actions at time $t+1$:
\begin{equation}
\begin{split}
IC^u&=I(m_t^u; a_{t+1}^u)\\
&=\sum_{a_{t+1}^u\in\mathcal{A_P}} \sum_{m_t^u\in\mathcal{M}_{DL}}{p(m_t^u, a_{t+1}^u) \log{ \frac{p(m_t^u, a_{t+1}^u)}{p(m_t^u) p(a_{t+1}^u)}}}.
\end{split}
\end{equation}
The marginal probabilities $p(m_t^u)$ and $p(a_{t+1}^u)$, and the joint probabilities $p(m_t^u, a_{t+1}^u)$ can be obtained by averaging \emph{\gls{dl} message} and \emph{channel access action} occurrences in simulation episodes after training.

\begin{figure}
\centerline{\includegraphics[width=\linewidth]{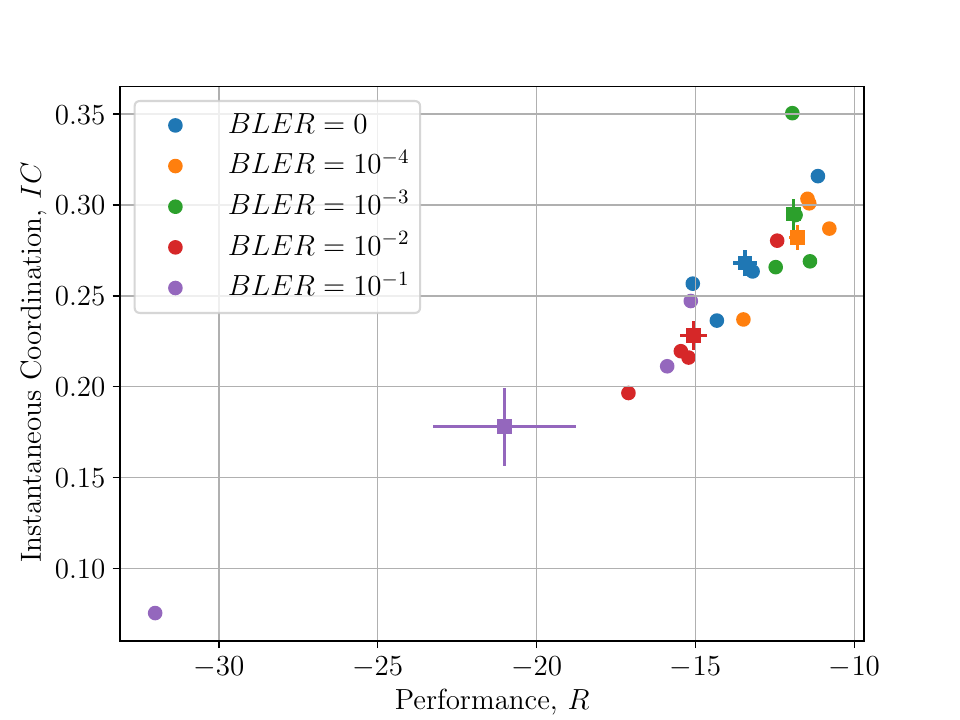}}
\caption{Relationship between the Instantaneous Coordination ($IC$) and performance $R$ across various training sessions. $IC$ is averaged across all \glspl{ue}. The length of the vertical error bars equal the standard error of the mean $IC$, while the length of the horizontal error bars equal the standard error of the mean performance (calculated over $4$ independent training sessions). The environment was configured with $|U|=2$, $P=2$, $t_{max}=32$ and empty start buffer. \gls{mac} learners were trained with a learning rate of $0.3$ and an internal memory length of $N=4$ for $N_{tr}=3M$ episodes. $\epsilon$-greedy exploration with $\epsilon$ decay from 1 to 0.01 was used.}
\label{fig:IC_vs_r}
\end{figure}

Fig. \ref{fig:IC_vs_r} shows that performance $R$ and Instantaneous Coordination $IC$ are clearly correlated (i.e., higher performance is concurrently achieved with higher $IC$).
The Pearson correlation coefficient $\rho_{IC,R}=0.91$ can be calculated as:
\begin{equation}
\rho_{IC,R}=\frac{cov(IC, R)}{\sigma_{IC}\sigma_R}
\end{equation}
where $\sigma_{IC}$ and $\sigma_R$ are the standard deviations of the Instantaneous Coordination and performance samples respectively.
This suggests that high performance may only be achievable under high influence (i.e., high $IC$) from \gls{bs} onto \glspl{ue}.
The signaling vocabulary (see Fig. \ref{fig:MACProtocolArchitecture}) plays thus a major role in the performance a \gls{mac} protocol can achieve.

\subsection{Generalization}
How well can the learned signaling and channel access policy perform in conditions never seen during training?
This question is about the generalization capacity of the learning algorithm (i.e., its robustness against environment variations, see \cite{Packer2019}).
We address this by first training the \gls{mac} learner in a default environment.
We then confront the trained learners against environments that differ in a single hyperparameter from the default parameters.

\begin{figure}[t]
\centerline{\includegraphics[width=\linewidth]{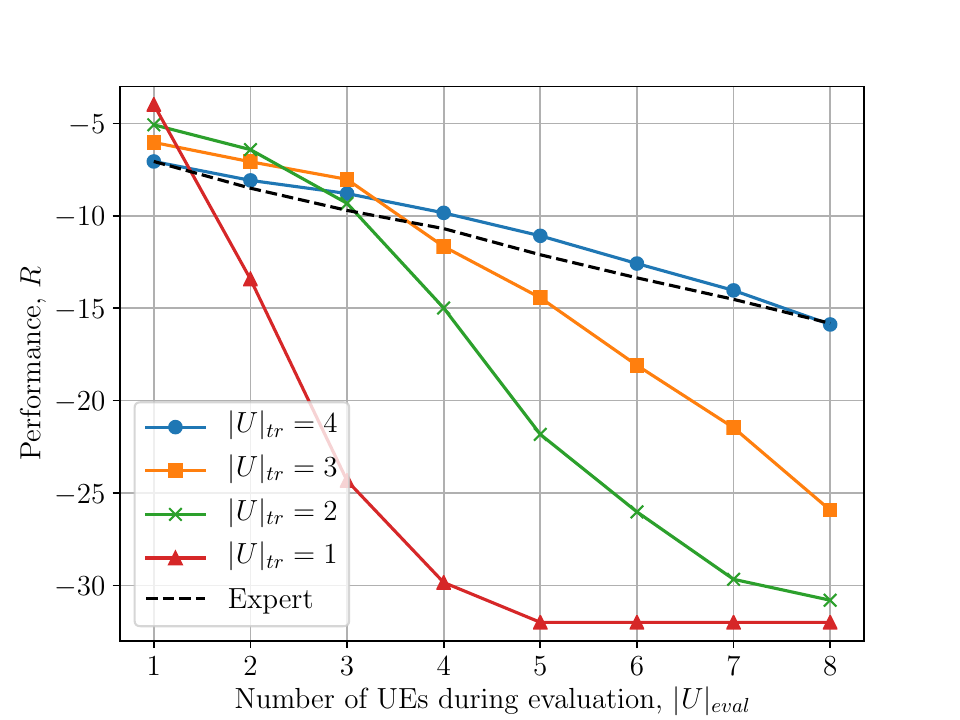}}
\caption{Best learner's performance (over $16$ independent training sessions). Training proceeds with $|U|_{tr}$ \gls{mac} learners but performance is evaluated with $|U|_{eval}$ \gls{mac} learners. The environment was configured with $P=1$, $BLER=10^{-4}$, $t_{max}=32$ and an empty start buffer. \gls{mac} learners were trained with a learning rate of $0.06$ and an internal memory length of $N=4$ for $N_{tr}=2^{24}=16M$ episodes. $\epsilon$-greedy exploration with $\epsilon$ decay from 1 to 0.01 was used.}
\label{fig:generalization_analysis_u}
\end{figure}

\begin{figure}
\centerline{\includegraphics[width=\linewidth]{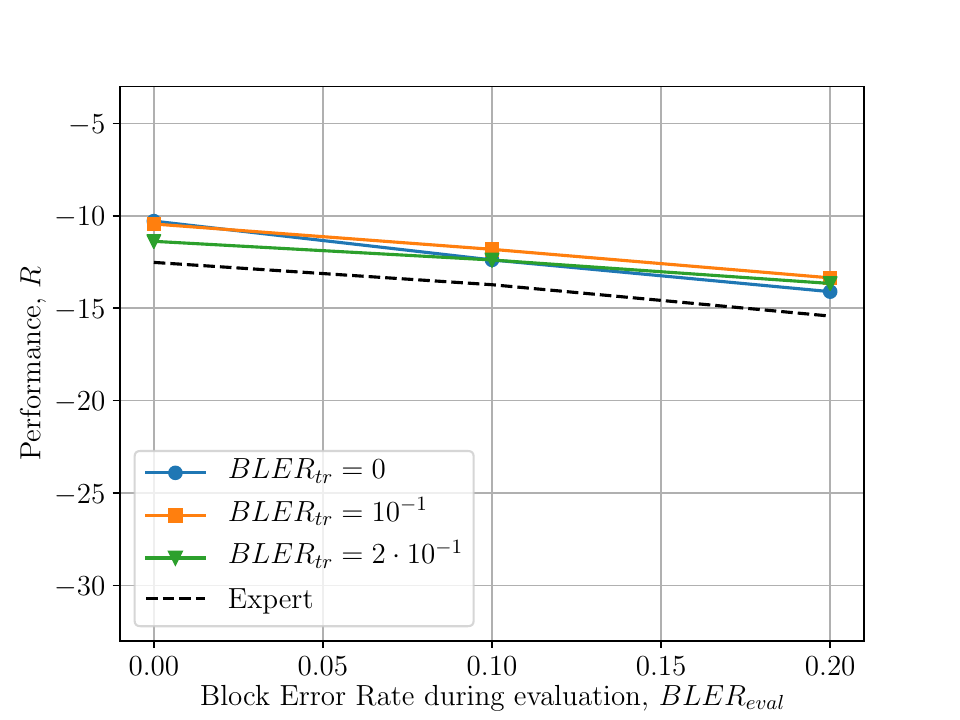}}
\caption{Best learner's performance (over $32$ independent training sessions). Training proceeds with $BLER_{tr}$ but performance is evaluated with $BLER_{eval}$. The environment was configured with $|U|=2$, $P=2$, $t_{max}=32$ and empty start buffer. \gls{mac} learners used a learning rate of $0.1$ and an internal memory length of $N=4$. $\epsilon$-greedy exploration with $\epsilon$ decay from 1 to 0.01 was used. Learners on the $BLER_{tr}=0, 10^{-1}, 10^{-2}$ environments were trained respectively for $N_{tr}=2^{23}, 2^{24}, 2^{25}$ ($\approx 8M, 16M, 32M$) episodes.}
\label{fig:generalization_analysis_bler}
\end{figure}

\begin{figure}
\centerline{\includegraphics[width=\linewidth]{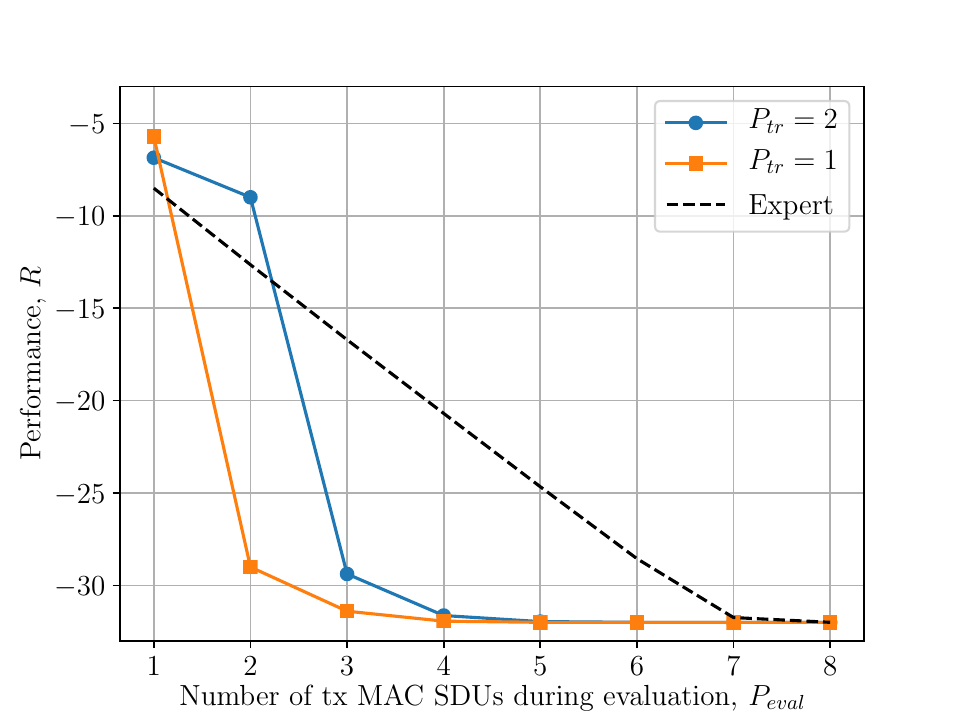}}
\caption{Best learner's performance and (over $4$ independent training sessions). Training proceeds with $P_{tr}$ \glspl{sdu}, but performance is evaluated with $P_{eval}$ \glspl{sdu}.} The environment was configured with $|U|=2$, $BLER=10^{-4}$, $t_{max}=32$ and an empty start buffer. \gls{mac} learners were trained with a learning rate of $0.06$ and an internal memory length of $N=4$ for $N_{tr}=2^{24}\approx16M$ episodes. $\epsilon$-greedy exploration with $\epsilon$ decay from 1 to 0.01 was used.
\label{fig:generalization_analysis_p}
\end{figure}

From Figs. \ref{fig:generalization_analysis_u} and \ref{fig:generalization_analysis_bler}, generalization seems robust across environment variations of \gls{bler} and $|U|$.
\gls{mac} learners can therefore be quickly trained in low-load scenarios with mild \gls{bler} conditions.
Then, these trained systems can still be expected to perform well in more challenging channels.
However, Figure~\ref{fig:generalization_analysis_p} shows that performance degrades rapidly with the number of \glspl{sdu} to be transmitted.
In this case, the trained learners have overfit to the number of \glspl{sdu} to be transmitted.
Training in environments with a larger number of \glspl{sdu} may alleviate this problem, although convergence in that case was elusive in our experiments.

\section{Conclusions}\label{sec:conclusions}
This paper has demonstrated that \gls{rl} agents can be jointly trained to learn a given simple \gls{mac} signaling and a wireless channel access policy.
It has done so using a purely tabular \gls{rl} approach in a simplified wireless environment.
It is expected that deep learning approaches may be able to scale these techniques to larger and more practical scenarios.
The paper has also shown that these agents can achieve the optimal performance when trained \mbox{\emph{tabula rasa}}, i.e., without any built-in knowledge of the target \gls{mac} signaling or channel access policy.

A main experimental observation of these experiments is that the trained agents understand but do not always comply with the DL signaling from the \gls{bs}.
This yields \gls{mac} protocols that can not be classified as contention-based or coordinated, but fluidly in between.
This is in line with \gls{5gnr}'s grant-free scheduling mechanism, although a major difference is that the protocols learned by our agents are not \gls{bs}-controlled.
A key advantage of protocols learned this way is that they can co-exist with the pre-existing human-designed protocol, while exploiting the optimizations uncovered during training.

Finally, this research has shown that the learned protocols depend on the deployment scenario.
This is one reason why they might yield higher performance than non-trainable protocols, whose signaling is static and independent of these parameters.

\subsection{Future work}
For the methodology presented here to be practical, a number of obstacles must be overcome.
The immediate next steps include a study on the sensitivity of the learned policies to the training environment.
This should respond to whether it is possible to train \gls{mac} learners in networks with a reduced number of \glspl{ue} and traffic volume, and then deploy them in larger networks.
A more detailed scalability analysis is also necessary (can these agents be trained in a reasonable amount of time in larger environments?).
The mirror problem of training a \gls{bs} to learn an expert \gls{ue} signaling may also illuminate the difficulties of protocol learning.

The ultimate goal is to jointly train \glspl{ue} and the \gls{bs} to emerge a fully new \gls{mac} protocol (i.e., to learn, not only the signaling and channel access policies, but also the signaling vocabulary of Fig. \ref{fig:MACProtocolArchitecture}).
Recent research (e.g., \cite{Jaques2018}, \cite{Lowe2019}) suggests this is hard when all agents (i.e., the \glspl{ue} and the \gls{bs}) begin with no previous protocol knowledge.
Consequently, scheduled training that alternates between supervised learning and self-play, as suggested in \cite{Lowe2020}, seems promising to emerge fully new protocols.

\bibliographystyle{IEEEtran}
\bibliography{IEEEabrv,MyLiterature}

\end{document}